\documentclass[aps,prl,twocolumn,showpacs,preprintnumbers]{revtex4}

\usepackage{graphicx}

\let\bm=\bibitem

\def\ft#1#2{{\textstyle{{\scriptstyle #1}\over {\scriptstyle #2}}}}
\def\fft#1#2{{#1 \over #2}}

\def\del{\partial}
\def\nn{\nonumber}
\def\sst#1{{\scriptscriptstyle #1}}
\def\0{{\sst{(0)}}}
\def\1{{\sst{(1)}}}
\def\2{{\sst{(2)}}}
\def\3{{\sst{(3)}}}
\def\4{{\sst{(4)}}}
\def\5{{\sst{(5)}}}
\def\6{{\sst{(6)}}}
\def\7{{\sst{(7)}}}
\def\8{{\sst{(8)}}}

\def\oneone{\rlap 1\mkern4mu{\rm l}}

\newcommand{\bea}{\begin{eqnarray}}
\newcommand{\eea}{\end{eqnarray}}
\newcommand{\be}{\begin{equation}}
\newcommand{\ee}{\end{equation}}

\begin{document}

\preprint{MIFP-05-13\ \ \ \ UPR-1125-T\ \ \ \ {\bf hep-th/0506029}}

\title{General Non-Extremal Rotating Black Holes in Minimal Five-Dimensional
   Gauged Supergravity}

\author{Z.-W. Chong$^1$, M. Cveti\v c$^2$, H. L\"u$^{1}$, 
C.N. Pope$^{1}$}
\affiliation{%
${}^1\!\!\!$ George P. \& Cynthia W.
Mitchell Institute for Fundamental Physics,
Texas A\&M University, College Station, TX 77843, USA\\
${}^2\!\!\!$ Department of Physics and Astronomy, University
of Pennsylvania, Philadelphia, PA 19104, USA
}%

\date{June 2, 2005}

\begin{abstract}

    We construct the general solution for non-extremal charged
rotating black holes in five-dimensional minimal gauged supergravity.
They are characterised by four non-trivial parameters, namely the
mass, the charge, and the two independent rotation parameters.  The
metrics in general describe regular rotating black holes, providing
the parameters lie in appropriate ranges so that naked singularities
and closed timelike curves (CTC's) are avoided. We calculate the
conserved energy, angular momenta and charge for the solutions, and
show how supersymmetric solutions arise in a BPS limit.  These have
naked CTC's in general, but for special choices of the parameters we
obtain new regular supersymmetric black holes or smooth topological
solitons.
 
\end{abstract}

\pacs{11.25.-w, 04.50.+h}
\maketitle

   The discovery of the remarkable AdS/CFT correspondence showed that
bulk properties of solutions in the five-dimensional gauged
supergravities that result from compactification of the type IIB
string are related to properties of strongly-coupled conformal field
theories on the four-dimensional boundary of five-dimensional anti-de
Sitter spacetime \cite{mald,guklpo,wit}.  It therefore becomes of
great importance to study the solutions of the five-dimensional gauged
supergravity theories.  One of the most important classes of such
solutions are those that describe black holes in five dimensions.  In
particular, it has been argued that the boundary conformal field
theory dual to rotating five-dimensional black holes should describe a
system in a four-dimensional rotating Einstein universe
\cite{hawhuntay}.

   The rotating five-dimensional solutions found in \cite{hawhuntay}
were neutral Kerr-(anti)-de Sitter black holes.  In order to be able
to make contact with supersymmetric BPS configurations, for which the
AdS/CFT correspondence is more solidly founded, it is of considerable
interest to generalise the neutral solutions to include electric
charge too.  In the analogous problem in ungauged supergravity, it is
straightforward to generate charged solutions from neutral ones, by
using the global symmetries of the ungauged supergravities as
solution-generating transformations.  By this means, the general
charged rotating black holes of five-dimensional ungauged supergravity
were obtained in \cite{cvetyoum}, starting from the neutral rotating
Ricci-flat black holes found in \cite{myerperr}.  For the solutions in
gauged supergravity there are no surviving global symmetries that can
be used to provide solution-generating transformations, and one has
little option but to resort to brute-force calculations, starting from
an appropriate ansatz, to construct the charged rotating solutions.
One way to simplify the problem is to specialise to the case where the
two independent rotation parameters of the generic five-dimensional
rotating black hole are set equal, since this reduces the problem from
cohomogeneity-2, with partial differential equations, to
cohomogeneity-1, with ordinary differential equations.  Supersymmetric
rotating black holes with two equal angular momenta were obtained in
\cite{gutreal}, and it was shown in \cite{gutreal} that the rotation
is necessary for the solution to be free of naked singularities and
CTC's.  The non-extremal charged rotating solutions of gauged
five-dimensional supergravity with equal rotation parameters were
constructed in \cite{d5gauge1,d5gauge2}.  Recently, some special cases
involving unequal rotation parameters were also constructed, in
\cite{d5gauge3}.  However, these latter arose as solutions of ${\cal
N}=2$ gauged supergravity coupled to two vector multiplets, with a
specific relation between the three electric charges, and did not, in
general, admit a specialisation to solutions of pure minimal ${\cal
N}=2$ gauged supergravity.  The purpose of this letter is to present
the general solution for charged rotating non-extremal black holes in
minimal five-dimensional gauged supergravity, with independent
rotation parameters in the two orthogonal 2-planes.

   We have found the general solution for charged rotating black holes
in five-dimensional minimal gauged supergravity, with unequal angular
momenta, by a process involving a considerable amount of trial and
error, followed by an explicit verification that the equations of
motion are satisfied.  In doing this, we have been guided by the
previously-obtained special case found in \cite{d5gauge1}, where the
two angular momenta were set equal, and the general charged rotating
solutions in ungauged minimal supergravity, which are contained within
the results in \cite{cvetyoum}.  In this letter we begin by presenting
our new solutions, and then we calculate the conserved angular momenta
and electric charge.  By integrating the first law of thermodynamics,
we also obtain the conserved mass, or energy, of the solutions.  By
considering the conditions under which the anticommutator of
supercharges in the AdS superalgebra has zero eigenvalues, we then
show how a BPS limit of our general non-extremal solutions gives rise
to new supersymmetric configurations.  These include new
supersymmetric rotating black holes, with two
independently-specifiable angular momenta, and new topological
solitons that are non-singular on complete manifolds.

   In terms of Boyer-Lindquist type coordinates $x^\mu= (t, r, \theta,
\phi,\psi)$ that are asymptotically static (i.e. the coordinate frame
is non-rotating at infinity), we find that the metric and gauge
potential for our new rotating solutions can be expressed as
\bea
ds^2 &=& -\fft{\Delta_\theta\, [(1+g^2 r^2)\rho^2 dt + 2q \nu]
\, dt}{\Xi_a\, \Xi_b \, \rho^2} + \fft{2q\, \nu\omega}{\rho^2}\nn\\
&&+ \fft{f}{\rho^4}\Big(\fft{\Delta_\theta \, dt}{\Xi_a\Xi_b} -
\omega\Big)^2 + \fft{\rho^2 dr^2}{\Delta_r} +
\fft{\rho^2 d\theta^2}{\Delta_\theta}\nn\\
&& + \fft{r^2+a^2}{\Xi_a}\sin^2\theta d\phi^2 + 
      \fft{r^2+b^2}{\Xi_b} \cos^2\theta d\psi^2\,,\label{5met}\\
A &=& \fft{\sqrt3 q}{\rho^2}\,
         \Big(\fft{\Delta_\theta\, dt}{\Xi_a\, \Xi_b} 
       - \omega\Big)\,,\label{gaugepot}
\eea
where
\bea
\nu &=& b\sin^2\theta d\phi + a\cos^2\theta d\psi\,,\nn\\
\omega &=& a\sin^2\theta \fft{d\phi}{\Xi_a} + 
              b\cos^2\theta \fft{d\psi}{\Xi_b}\,,\nn\\
\Delta_\theta &=& 1 - a^2 g^2 \cos^2\theta -
b^2 g^2 \sin^2\theta\,,\nn\\
\Delta_r &=& \fft{(r^2+a^2)(r^2+b^2)(1+g^2 r^2) + q^2 +2ab q}{r^2} - 2m 
\,,\nn\\
\rho^2 &=& r^2 + a^2 \cos^2\theta + b^2 \sin^2\theta\,,\nn\\
\Xi_a &=&1-a^2 g^2\,,\quad \Xi_b = 1-b^2 g^2\,,\nn\\
f&=& 2 m \rho^2 - q^2 + 2 a b q g^2 \rho^2\,.
\eea
A straightforward calculation shows that these configurations solve
the equations of motion of minimal gauged five-dimensional
supergravity, which follow from the Lagrangian
\be
{\cal L} = (R+ 12g^2)\, {*\oneone} - \ft12 {*F}\wedge F + \fft1{3\sqrt3} 
      F\wedge F\wedge A\,,
\ee
where $F=dA$, and $g$ is assumed to be positive, without loss of
generality.

   For some purposes, it is useful to note that the non-vanishing
metric components are given by
\bea
g_{00}\!\!\!&=&\!\!\! -\fft{\Delta_\theta (1+g^2 r^2 )}{\Xi_a\, \Xi_b} + 
         \fft{\Delta_\theta^2 (2m \rho^2 -q^2 + 2 a b q g^2\rho^2)}{
         \rho^4 \Xi_a^2\, \Xi_b^2}\,,\nn\\
g_{03}\!\!\! &=&\!\!\! -\fft{\Delta_\theta\, [a (2m \rho^2-q^2) + 
     b q\rho^2(1+a^2 g^2)]\sin^2\theta}{\rho^4 \Xi_a^2\, \Xi_b}\,,\nn\\
g_{04}\!\!\! &=&\!\!\! -\fft{\Delta_\theta\,[b (2m \rho^2-q^2) + 
     a q\rho^2(1+b^2 g^2)]\cos^2\theta }{\rho^4 \Xi_b^2\, \Xi_a}\,,\nn\\
g_{33}\!\!\! &=&\!\!\! \fft{(r^2+a^2)\sin^2\theta}{\Xi_a} + 
                \fft{ a[a(2m\rho^2 -q^2) + 2 bq \rho^2]\sin^4\theta}{
              \rho^4\Xi_a^2}\,,\nn\\
g_{44}\!\!\! &=&\!\!\! \fft{(r^2+b^2)\cos^2\theta}{\Xi_b} + 
                \fft{ b[b(2m\rho^2 -q^2) + 2 aq \rho^2]\cos^4\theta}{
              \rho^4\Xi_b^2}\,,\nn\\
g_{34}\!\!\!&=&\!\!\!
\fft{[ ab(2m \rho^2 -q^2) + (a^2+b^2) q \rho^2]\, \sin^2\theta\,
             \cos^2\theta}{\rho^4\Xi_a\, \Xi_b}\,,\nn\\
g_{11}\!\!\! &=&\!\!\! \fft{\rho^2}{\Delta_r}\,,\qquad g_{22}= 
                \fft{\rho^2}{\Delta_\theta}\,.
\eea

   The Killing vector 
\be
\ell = \fft{\del}{\del t} + \Omega_a\, \fft{\del}{\del \phi} 
    + \Omega_b \, \fft{\del}{\del \psi}
\ee
becomes null on the outer Killing horizon at $r=r_+$, the largest
positive root of $\Delta_r=0$, where the angular velocities on the
horizon are given by
\bea
\Omega_a &=& \fft{a(r_+^2+ b^2)(1+g^2 r_+^2) + b q}{
               (r_+^2+a^2)(r_+^2+b^2)  + ab q}\,,\nn\\
\Omega_b &=& \fft{b(r_+^2+ a^2)(1+g^2 r_+^2) + a q}{
               (r_+^2+a^2)(r_+^2+b^2)  + ab q}\,.
\eea
One can then easily evaluate the surface gravity
\be
\kappa = \fft{r_+^4[(1+ g^2(2r_+^2 + a^2+b^2)] -(ab + q)^2}{
         r_+\, [(r_+^2+a^2)(r_+^2+b^2) + abq]}\,,
\ee
and hence the Hawking temperature $T=\kappa/(2\pi)$.  The entropy
is given by
\be
S=\fft{\pi^2 [(r_+^2 +a^2)(r_+^2 + b^2) +a b q]}{2\Xi_a \Xi_b r_+}
\,.
\ee

   The angular momenta can be evaluated from the Komar integrals
$J = 1/(16\pi)\int_{S^3}  {*dK}$, where $K=\del/\del\phi$ or
$K=\del/\del\psi$, yielding
\bea
J_a &=& \fft{\pi[2am + qb(1+a^2 g^2) ]}{4 \Xi_a^2\, \Xi_b}\,,\nn\\
J_b &=& \fft{\pi[2bm + qa(1+b^2 g^2) ]}{4 \Xi_b^2\, \Xi_a}\,.
\eea
The electric charge follows from the Gaussian integral 
$Q=1/(16\pi)\int_{S^3}( {*F} -F\wedge A/\sqrt3)$,  yielding 
\be
Q = \fft{\sqrt3\, \pi q}{4 \Xi_a\, \Xi_b}\,.
\ee
Using the technique introduced in \cite{gibperpop}, the easiest way
to calculate the conserved mass, or energy, is to integrate the first
law of thermodynamics $dE=TdS + \Omega_a dJ_a + \Omega_b dJ_b +
\Phi dQ$, where $\Phi= \ell^\mu A_\mu$ is the electrostatic potential 
on the horizon.  Doing this, we find
\be
E= \fft{m\pi(2\Xi_a+2\Xi_b - \Xi_a\, \Xi_b) +
2\pi q a b g^2(\Xi_a+\Xi_b)}{4 \Xi_a^2\, \Xi_b^2}\,.
\ee

   The BPS limit can be found by looking at the eigenvalues of the
Bogomol'nyi matrix coming from the anticommutators of the
supercharges, as discussed in \cite{cvgilupo}.  Thus we have BPS
solutions if
\be
E - g J_a - g J_b - \sqrt3\, Q =0\,.
\ee
From the expressions derived above for $(E, J_a, J_b, Q)$, we find that
the BPS limit is achieved if
\be
q = \fft{m}{1 + (a+b)g}\,.\label{susycon}
\ee
The supersymmetry of the solutions in this limit can be confirmed by 
calculating the norm of the Killing vector
\be
K_+ \equiv \fft{\del}{\del t} + g\, \fft{\del}{\del\phi}
    + g \, \fft{\del}{\del\psi}\,,
\ee
which, as discussed in \cite{cvgilupo}, arises as the square of the
Killing spinor $\eta$, in the sense that
$K_+^\mu=\bar\eta\gamma^\mu\eta$.  We find that its norm is given by
\be
K_+^2 = -\fft{[h - m(1+ag \cos^2\theta + bg\sin^2\theta)]^2}{h^2}\,,
\label{kp}
\ee
where
\be
h= (1+ag)(1+bg)[1+(a+b)g]\rho^2\,.
\ee
Thus indeed the norm of $K_+$ is, as it should be since it has a
spinorial square root, manifestly negative definite.  The fraction of
supersymmetry preserved is in general $\ft14$, except when $a=-b$, in
which case, the preserved supersymmetry is doubled to become $\ft12$.
The latter solution was previously obtained in \cite{ks}.

    We now discuss the global structure of the rotating AdS$_5$ black
hole.  To do this, we first note that the metric can be expressed as
\bea
ds^2&=& -\fft{\Delta_r\Delta_\theta r^2\sin^2 2\theta}{4\Xi_a^2
\Xi_b^2 B_\phi B_\psi} dt^2 + \rho^2 (\fft{dr^2}{\Delta_r} +
\fft{d\theta^2}{\Delta_\theta})  \label{met2}\\
&&+ B_\psi (d\psi + v_1 d\phi +
v_2 dt)^2 + B_\phi (d\phi + v_3 dt)^2\,,\nn
\eea
where the functions $B_\phi$, $B_\psi$, $v_1$, $v_2$ and $v_3$ can be
straightforwardly found by comparing (\ref{met2}) with the metric in
(\ref{5met}).  The absence of naked closed timelike curves (CTC's)
requires that $B_\phi$ and $B_\psi$ be non-negative outside the
horizon.  We shall focus on the discussion of supersymmetric
solutions, satisfying the condition (\ref{susycon}).  It can be seen
from (\ref{kp}) that the identity
\bea
\!\!\!&&\!\!\!-\fft{\Delta_r\Delta_\theta 
r^2\sin^2 2\theta}{4\Xi_a^2 \Xi_b^2 B_\phi B_\psi} +
B_\psi (v_2 + g + g v_1)^2 +
B_\phi (v_3 + g)^2\nn\\
\!\!\!&&\!\!\!=-\fft{[h - m(1+ag \cos^2\theta +
bg\sin^2\theta)]^2}{h^2}   \label{identity}
\eea
holds. It follows that in general, at the Killing horizon where
$\Delta_r=0$, we have $B_\phi\cdot B_\psi<0$, implying the existence
of naked CTC's.  There are two special cases where naked CTC's can be
avoided, leading to either supersymmetric black holes or topological
solitons:

\bigskip
{\noindent \it Supersymmetric black holes}:  The first way to
avoid naked CTC's is if the right-hand side of (\ref{identity})
vanishes on the Killing horizon.  This occurs when the 
parameters in the supersymmetric solutions satisfy the further
restriction
\be
g m = (a + b) (1 + a g) (1 + b g) (1 + a g + b g)\,.
\ee
Remarkably, when this extra condition is satisfied,
the function $\Delta_r$ has a double root; $\Delta_r$ is now given by
\be
\Delta_r = r^{-2} (r^2-r_0^2)^2 [g^2 r^2 + (1 + ag + bg)^2]\,,
\ee
where $r_0^2 = g^{-1} (a + b + a b g)$. At the Killing horizon
$r=r_0$, we find that the determinant of the metric in the $(\theta,
\phi, \psi)$ directions is given by
\be
\det g_{(\theta,\phi,\psi)}
= \fft{(a+b)^2 (a + b + a b g) \sin^2 2\theta }{4 g^3 (1 - ag)^2
(1-bg)^2}
\ee
This implies that naked CTC's are avoided if the remaining free
parameters $a$ and $b$ satisfy the inequality 
\be
a + b + a b g > 0\,.
\ee
The Killing horizon $r=r_0$ is then the event horizon of a
well-defined supersymmetric black hole that is regular on and outside
the event horizon.  The occurrence of the double-root of $\Delta_r$ at
$r=r_0$ implies that the black hole has zero temperature.  The various
conserved and thermodynamic quantities for these new supersymmetric
black holes are given by
\bea
E&=&\fft{\pi (a + b)}{4g (1 -a g)^2 (1-bg)^2}\Big( (1-ag)(1-bg)\nn\\
&& \qquad +(1 +ag)(1+bg) (2-ag-bg)\Big)
\,,\nn\\
S&=&\fft{\pi^2(a +b) \sqrt{a + b + a b g}}{2 g^{3/2} (1-ag)
(1- bg)}\,,\nn\\
J_a&=&\fft{\pi (a +b) (2a +b + a bg)}{4g (1-a g)^2 (1-bg)}\,,\nn\\
J_b&=&\fft{\pi (a +b) (a + 2b + abg)}{4g (1-a g)(1-bg)^2}\,,\nn\\
Q &=& -\fft{\pi\sqrt3 (a + b)}{4 g (1 -a g)(1 - b g)}\,.\label{susytherm}
\eea
Note that supersymmetric black holes cannot arise when $a=-b$.  For
$a=b$, our new supersymmetric black hole solutions, which for general
$a$ and $b$ have cohomogeneity 2, become cohomogeneity 1; these
special cases were previously obtained in \cite{gutreal}.

\bigskip
{\noindent \it Topological solitons}:   The second way to
avoid naked CTC's is if $B_\phi=0$ at $r=r_0$.  This can happen
when the free parameters in the general supersymmetric solutions
obey the further restriction 
\bea
m&=&-(1 + ag)(1+bg)(1 + ag + bg)\nn\\
&&\qquad \times (2a +b + abg)(a + 2b + abg)\,.
\eea
Now $r_0$, the outer root of $\Delta_r$, is given by
\be
r_0^2 = - (a + b + a b g)^2\,.
\ee
Defining a new radial coordinate $R=r^2- r_0^2$, we find that the
metric describes a smooth topological soliton, with $R$ running from 0
to $\infty$.  The requirement of the absence of a conical singularity
when $B_\phi$ vanishes at $R=0$ implies the quantisation condition
\be
\fft{(a + b + a b g) (3 + 5 a g + 5bg + 3abg^2)}{(1-a g)
(a + 2b + abg)}=1
\,.
\ee
In the cohomogeneity-1 special cases $a=b$ or $a=-b$, these toplogical
solitons are encompassed within the soliton solutions obtained in
\cite{cvgilupo}.

       Aside from the above two possibilities, the supersymmetric
solutions in general have naked CTC's.  As in the examples discussed
in \cite{cvgilupo,d5gauge3}, a conical singularity at the Killing
horizon can be avoided by periodically identifying the asymptotic time
coordinate $t$ with an appropriate period.  However, if the Killing
horizon is associated with a double root of $\Delta_r$, then such an
identification is unnecessary, analogous to the ungauged rotating
solution obtained in \cite{bmpv}.  The geodesic analysis of analogous
time machines can be found in \cite{gibher,cks}.

   In the general case where the charged rotating metrics that we
have found are non-extremal, they describe regular black holes provided
the parameters lie in appropriate ranges that are easily determinable
using the same techniques we have used above for analysing the BPS
limits.

   As discussed in \cite{hawhuntay}, rotating black hole solutions in
five-dimensional gauged supergravity provide backgrounds whose AdS/CFT
duals describe four-dimensional field theories in the rotating
Einstein universe on the boundary of anti-de Sitter spacetime.  With
the general solutions in minimal gauged supergravity that we have now
found, this aspect of the AdS/CFT correspondence can be studied in a
framework that also allows one to take a BPS or near-BPS limit, where
the mapping from the bulk to the boundary is better controlled.  In
particular, it is of great interest to provide the microscopic
interpretation from the boundary CFT for the entropy (\ref{susytherm})
of the supersymmetric black holes with two general rotations.  We plan
to report further on these considerations in forthcoming work.

\bigskip
\noindent{\bf Acknowledgements}:

We thank Gary Gibbons for useful discussions.  C.N.P. thanks the
Relativity and Cosmology group in DAMTP, Cambridge, for hospitality
during the course of this work.  Research supported in part by DOE
grants DE-FG02-95ER40893 and DE-FG03-95ER40917, NSF grant INTO3-24081,
and (M.C.) the University of Pennsylvania Research Foundation Award
and the Fay R. and Eugene L.Langberg Chair.

\end{document}